\documentclass[fleqn,usenatbib]{mnras}

\usepackage{newtxtext,newtxmath}

\usepackage[T1]{fontenc}

\DeclareRobustCommand{\VAN}[3]{#2}
\let\VANthebibliography\thebibliography
\def\thebibliography{\DeclareRobustCommand{\VAN}[3]{##3}\VANthebibliography}

\usepackage{graphicx}	
\usepackage{amsmath}	
\usepackage{amssymb}	
\usepackage{lipsum}     
\usepackage{siunitx}    
\usepackage{bm}         
\usepackage{orcidlink}

\title[Obliquities from gravito-turbulence]{Primordial Obliquities of Brown Dwarfs and Super-Jupiters from Fragmenting Gravito-Turbulent Discs}

\author[Jennings \& Chiang]{R. Michael Jennings$^{1}$\thanks{E-mail: robertmjenningsjr@berkeley.edu}\orcidlink{0000-0002-3959-6572},
Eugene Chiang$^{1,2}\orcidlink{0000-0002-6246-2310}$
\\
$^{1}$Department of Astronomy, University of California, Berkeley, Berkeley, CA 94720\\
$^{2}$Department of Earth and Planetary Science, University of California, Berkeley, Berkeley, CA 94720\\
}

\pubyear{2021}

\begin{document}
\label{firstpage}
\pagerange{\pageref{firstpage}--\pageref{lastpage}}
\maketitle

\begin{abstract}
Super-Jupiters, brown dwarfs, and stars can form from the collapse of self-gravitating discs. Such discs are turbulent, with flocculent spiral arms accelerating gas to transonic speeds horizontally and vertically. Objects that fragment from gravito-turbulent discs should spin with a wide range of directions, reflecting the random orientations of their parent eddies. We show by direct numerical simulation that obliquities of newly collapsed fragments 
can range up to 45$^\circ$.
Subsequent collisions between fragments can further alter the obliquity distribution, up to 90$^\circ$ or down to near-zero. The large obliquities of newly discovered super-Jupiters on
wide orbits around young stars may be gravito-turbulent in origin. Obliquely spinning fragments are born on orbits that may be inclined relative to their parent discs by up to 20$^\circ$, and gravitationally stir leftover material to many times the pre-fragmentation disc thickness.
\end{abstract}




\section{Introduction} \label{sec:intro}

Planets that form from unwarped razor-thin circumstellar discs have no choice but to have zero obliquity at birth, their spin axes and orbit normals perfectly aligned.\footnote{The vector 
spin angular frequency of a fluid
element equals half its
vorticity (e.g. \citealt{Shapiro1961}). If velocities are purely horizontal,
the vector vorticity is vertical. In Rayleigh-stable discs where the specific angular momentum increases radially outward, the vorticity points in the same direction as the orbital angular momentum, i.e. fluid elements spin prograde.} Small obliquities $\ll 1$ radian also result from finite-thickness discs if the vertical velocities of accreted material sum to near-zero. The larger the spatial scales over which an object accumulates its mass, the greater the degree of cancellation.

Planets derived from material on small scales, in discs with significant vertical motions, may form with large obliquities. A 3D turbulent disc is filled with eddies whirling in random directions. A tilted eddy that gravitationally collapses on itself may retain its vector spin angular momentum to become an obliquely rotating protoplanet. Density fluctuations in the disc may also torque spins. 

Turbulence and gravitational fragmentation go hand in hand. The faster self-gravitating discs cool, the more they are roiled by spiral density waves. A gravito-turbulent disc in 3D exhibits transonic radial and vertical motions extending over many vertical scale heights \citep{Shi2014}. If the disc's cooling time becomes shorter than about an orbit time, the disc fragments \citep{Gammie2001}. On larger molecular cloud scales, turbulence leads to random fluctuations in the vector angular momentum of material that accretes to form a star and its attendant disc, potentially yielding a non-zero stellar obliquity \citep[e.g.][]{Bate2010,Thies2011,Fielding2015}.

Self-gravitating discs are
thought to spawn brown dwarf and super-Jupiter companions to stars on wide orbits. The top-heavy mass function of brown dwarf secondaries; their tendency to be found at large distances from their host stars; and their lack of preference for host stellar type all
point to top-down formation by disc fragmentation (e.g. \citealt{Nielsen2019};
see also \citealt{Kratter2016}).
Recent observations of 2M0122b---a 12--27 Jupiter mass object orbiting an M dwarf at a projected separation of 50 au---indicate its obliquity may be large,
although the data for this system are
not conclusive because of sky-projection
degeneracies 
\citep{Bryan2020}. 
The case
of HD 106906 is more secure; the F star binary hosts, at separations
of hundreds of au's, a $\sim$12 Jupiter mass companion viewed nearly pole-on,
and a debris disc 
viewed nearly edge-on (Bryan et al.~2021, in press).
Large obliquities of 
super-Jupiters on wide orbits
suggest fragmentation of a gravito-turbulent
disc.

We aim here to provide
further evidence supporting
this interpretation. 
We present 3D numerical fluid simulations
of secularly cooling, gravito-turbulent discs and
measure the obliquities of collapsed
fragments. A shearing box is employed to resolve the turbulent structure of a local patch of the disc. The 
drawback of the shearing box is 
that it cannot track the evolution
of fragment orbits; orbital radius and by extension orbital angular momentum are not defined in a shearing box formalism  which neglects radial curvature terms. Thus for obliquity, which measures the angle between the spin and orbital angular momentum vectors, we can only track the spin vector; the orbit normal is assumed fixed along the vertical axis of the box.
This shortcoming is removed in a global calculation at the cost of spatial resolution. \citet{Hall2017} presented global 3D smooth-particle-hydrodynamics simulations of self-gravitating discs, apparently finding all bound fragments to form with obliquities $\lesssim 10^\circ$. Collisions, and mass and angular momentum transfer between fragments led to four fragments among the forty recorded spinning retrograde with obliquities $> 160^\circ$. One fragment was plotted to have an obliquity of $35^\circ$, but its dynamical history was not discussed. We offer here a complementary calculation that
more finely resolves the 3D velocity fields in self-gravitating discs, in particular the vertical flows that can lead to oblique rotators.

The rest of this paper is organized
as follows. Section~\ref{sec:code} describes
our numerical code and setup, 
section~\ref{sec:results} presents our results
on obliquities and other fragment properties, and section~\ref{sec:sum} provides
a summary and outlook.

\section{Code and Setup} \label{sec:code}

Our computational tools and procedures are
similar to those of \citet{Shi2014}, and we provide here just a brief description highlighting differences.
We use the code {\tt Athena}
to solve
the hydrodynamic equations in a 3D shearing box,
including vertical tidal gravity from the central
star \citep{Stone2008}. We adopt the van Leer 
integrator \citep{VanLeer2003,Stone2009a}; 
a piecewise linear spatial reconstruction in the 
primitive variables; the Harten–Lax–van Leer contact 
(HLLC) Riemann solver; and orbital advection 
algorithms that speed runtimes and 
improve conservation \citep{Masset2000,Johnson2008,Stone2010}. Disc self-gravity is included by solving Poisson's equation using fast Fourier transforms 
\citep{Koyama2009,Kim2011}.
Boundary conditions are shearing-periodic in radial displacement $x$, and periodic in azimuth $y$. In vertical height $z$, vacuum boundary conditions are adopted for
the self-gravitational potential $\Phi$, and
reflecting boundary conditions for
the other hydrodynamic flow variables.
This last choice differs from \citet{Shi2014}
who used vertical outflow boundary conditions.
We found that outflow conditions resulted in significant mass loss from the simulation domain after
fragments form and gravitationally stir the remaining disc material (see section \ref{sec:vert}). Mass loss was not an issue for \citet{Shi2014} whose focus was on turbulence in non-fragmenting discs. With reflecting vertical boundary conditions for density and velocity,
the mass in the box is conserved to a few percent.

\begin{figure}
    \centering
    \includegraphics[width=\columnwidth]{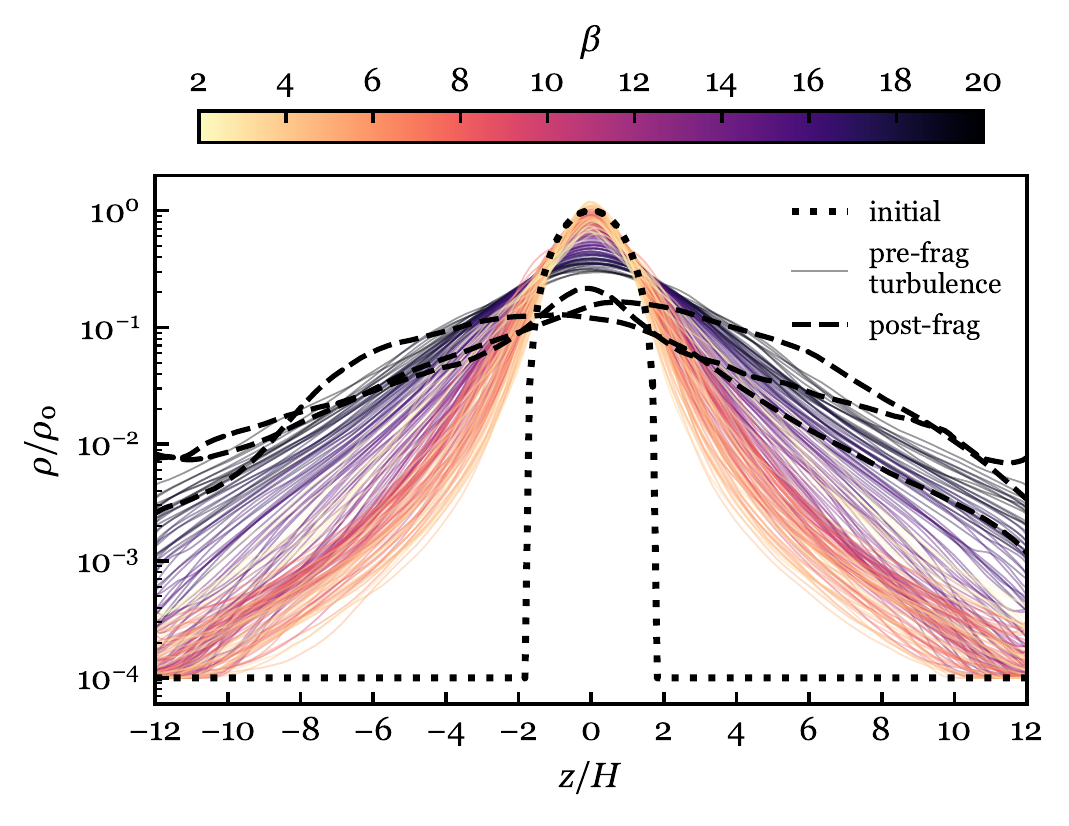}
    \caption{Vertical density profiles of the background circumstellar
      (CS) disc in our high-resolution (8 cells per $H$), $\beta_0=20$
      run. The initial condition (vertical hydrostatic equilibrium) is
      plotted as a dotted line. When the disc becomes
      gravito-turbulent, the disc thickens. Thin colored lines are
      horizontally averaged snapshots plotted before the disc
      fragments; as $\beta = \Omega t_{\rm cool}$ decreases from 20 to
      2 and cooling accelerates, the disc flattens.
    The $\beta = 2$ curve describes the vertical density
  profile just before the fragment forms. At this stage the disc is
  relatively thin and the density profile within $z = \pm 2H$
  resembles the hydrostatic initial condition.
    Once a fragment forms the disc puffs back up, and 
its midplane oscillates about $z=0$,
    as seen by the dashed lines representing horizontally averaged snapshots with the fragment removed. The stirring is caused by the gravity of the fragment. 
    }
    \label{fig:zprofile}
\end{figure}

Initial conditions at time $t=0$ are of a horizontally uniform disc
with a vertical density profile in 
hydrostatic equilibrium including self-gravity:
\begin{equation} \label{eq:HE}
    \frac{1}{\rho}\frac{dP}{dz} = -\Omega^2z - 4\pi G\int^z_0 \rho(z')dz' 
\end{equation}
for density $\rho$, pressure $P$, fixed orbital angular
frequency $\Omega$, and gravitational constant $G$.
Given an initial surface density $\Sigma_0$ and initial midplane
density $\rho_0$, we define a characteristic initial disc half-thickness $H \equiv \Sigma_0/(2\rho_0)$ over which most of the disc mass is contained. The gas is assumed to obey a polytrope $P = K \rho^{\gamma}$ initially, with an initial midplane sound speed
$c_{\rm s0} = \sqrt{\gamma K \rho_0^{\gamma-1}}$.
Upon setting $\gamma=5/3$ and $Q_0 \equiv c_{\rm s0} \Omega / (\pi G \Sigma_0) = 1$, equation (\ref{eq:HE}) may be solved for the initial density profile $\rho(z)$ \citep{Shi2013,Shi2014}. The solution is plotted as a dotted curve in Figure \ref{fig:zprofile}. Analytically, $\rho$ vanishes at a finite height $z \simeq \pm 2H$. To avoid having to resolve steep density and pressure
gradients near the vertical boundaries of the box, we introduce a density floor of $\rho_{\rm min} = 10^{-4} \rho_0$. We work in code units for which $\rho_0 = \Omega = H = 1$; it follows
that $\Sigma_0 = 2$, $c_{\rm s0} = 2.126$,
$K = 2.712$, and $G = 0.3384$.
To seed gravitational instability we introduce (at $t=0$ only) random cell-to-cell velocity perturbations with a maximum amplitude of either 
$0.02$ or $0.2$$c_{\rm s0}$ for $|z| < 2H$; we have verified in a few cases that our results do not depend on the exact choice of initial perturbation.

Our fiducial box size is $\{L_x, L_y, L_z\} = \{16H, 64H, 24H \}$. We want our
box to be large enough to avoid much disc material, and most
especially bound fragments, crashing into the reflecting box lids, and
to accommodate as many turbulent sub-structures (potential fragments)
as possible. If we say that our simulation is intended to model a
portion of the disc at radius $a$ having a typical protoplanetary disc
aspect ratio $H/a =0.1$, then our box extends the full disc
circumference in azimuth, and $0.2a$--$1.8a$ in radius (in order-unity
violation of the shearing sheet formalism which assumes the
computational domain is less than the disc radius in size). 
The azimuthal length $L_y$ is large enough to accommodate
the long-wavelength modes responsible for maintaining steady
gravito-turbulence without ``bursty'' behaviour \citep{Booth2019}.
Our fiducial resolution is 4 cells per $H$;
we also experiment with high-resolution
runs having 8 cells per $H$.
\citet{Shi2014} used identical resolutions and
  found no appreciable
  difference between them; similar results were obtained by \citet{Booth2019} who tested up to 32 cells per $H$.

In our fiducial runs, cooling is effected by lowering the internal energy density $U = P/(\gamma-1)$ of every cell at a rate $\dot{U} = -U/t_{\rm cool}$ for cooling time $t_{\rm cool} \equiv \beta/\Omega$. 
In addition to the aforementioned floor on mass density $\rho$, we impose a floor on the energy density (and by extension pressure) of $U_{\rm min} = 10^{-4}$ in code units. We start 
runs with $\beta = \beta_0 \in \{ 10,20 \}$ for
times $t < 40\Omega^{-1}$; during this earliest phase of the simulation, the disc is in a quasi-steady, non-fragmenting, gravito-turbulent state.
To induce fragmentation, we then accelerate cooling by lowering $\beta$ from $\beta_0$ to a minimum value of 2 
over a linear ramp in time from $t = 40\Omega^{-1}$ to
$t = 120\Omega^{-1}$. Typically one or two fragments form when $\beta = 2$.

We also experiment with an alternative
cooling prescription introduced by \citet{Shi2014} to simulate radiative cooling in an optically thin medium with constant opacity:
\begin{equation} \label{eq:opt}
    \dot{U} = -\frac{1}{b(\gamma -1)}\frac{P^4}{\rho^3} \,.
\end{equation}
The cooling time $U/|\dot{U}|$ then scales as $[P/(\gamma-1)]/|\dot{U}| = b(\rho/P)^3 \propto b/T^3$ for coefficient $b$ and temperature $T$; this scaling is appropriate for gas whose energy density scales as $\rho T$ and whose volume emissivity scales as $\rho T^4$. We start with $b=b_0 \in \{500,600\}$ for the first
$40\Omega^{-1}$, and then accelerate cooling by ramping $b$ down at a rate of $db/dt = 100/(80\Omega^{-1})$.
Note that these optically thin cooling experiments are performed to explore an issue of principle, namely the sensitivity of obliquity outcomes to the cooling prescription, and not because we think massive, self-gravitating discs are necessarily optically thin in practice. Early in the evolution of a disc, when it is still being fed by a natal envelope, we suspect it is optically thick to its own cooling radiation.

\section{RESULTS} \label{sec:results}

\begin{figure*}
    \centering
    \includegraphics[width=\textwidth]{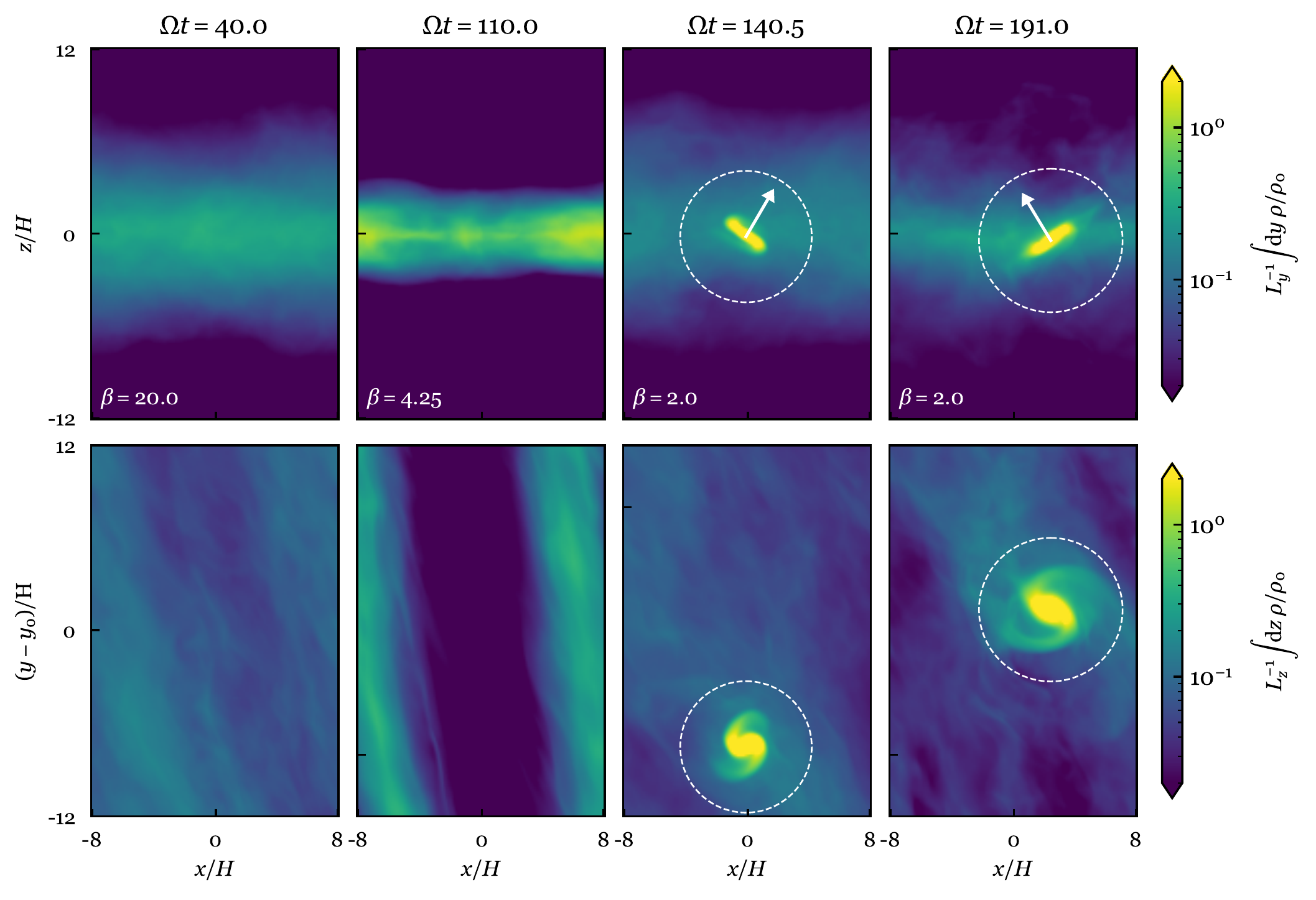}
    \caption{Column density snapshots  
    of our high resolution (8 cells per $H$), $\beta_0=20$ run. The top row, with snapshot times listed in the header and corresponding $\beta = \Omega t_{\rm cool}$ values annotated in the corners, shows edge-on views of the disc in the meridional $x$-$z$ plane, with density integrated in the azimuthal-$y$ direction. The bottom row shows face-on views of the disc, plotting surface density in $x$-$y$ plane (density integrated along the vertical $z$ direction) for subsections of the simulation domain. At early times, trailing spiral arms amplify.  
    Later, a single fragment forms whose spin angular momentum vector $\bm{J}$ points $\sim$35 degrees from the vertical. The white arrow of arbitrary length shows the $x$-$z$ projection of $\bm{J}$ and the dashed circle encloses the fragment's Hill sphere. The fragment is disc-shaped (top row), exhibits two tidal arms (bottom row), and has a spin axis which is directed in the positive $z$-direction (the spin is prograde; see Fig.~\ref{fig:contours}) and which precesses about the $z$-axis with angular frequency $\Omega$ in the retrograde direction. The precession causes the apparent orientation of the fragment (a.k.a.~circumplanetary disc) to change between the third and fourth columns and is a purely kinematic effect arising from the rotating frame.}
    \label{fig:proj}
\end{figure*}

\begin{figure*}
    \centering
    \includegraphics[width=0.8\textwidth]{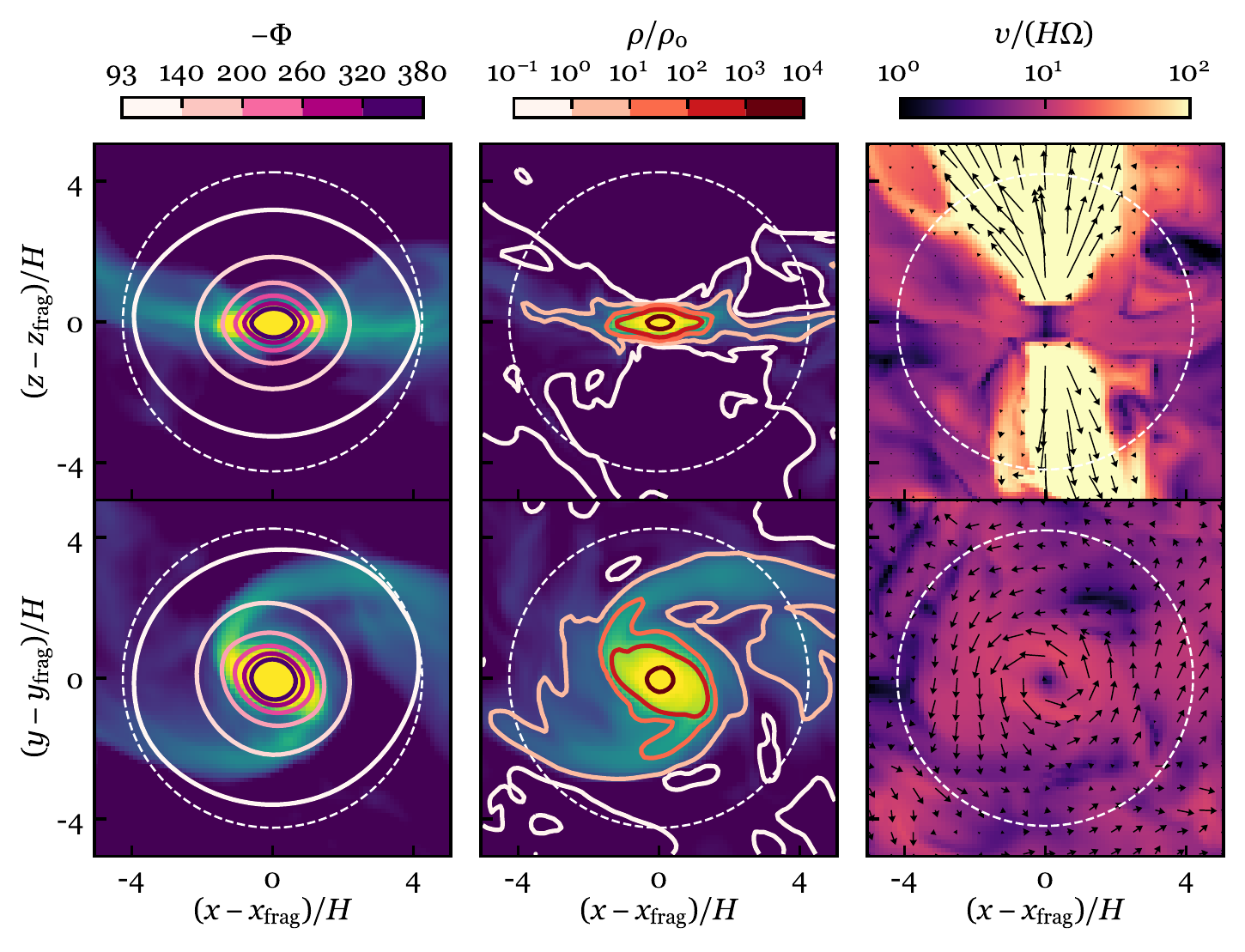}
    \caption{{\it Left:} Iso-contours of the self-gravitational potential $\Phi$, taken in an $x$-$z$ slice (at fixed $y$ passing through the fragment center of mass; top panel)
    and an $x$-$y$ slice (at fixed $z$ passing through the fragment center of mass; bottom panel), for the  high-resolution, $\beta_0=10$ run at $\Omega t = 112$. Color background traces column densities.
    {\it Center:} Volume density iso-contours for these same slices. Color background traces volume density.
    {\it Right:} Velocities for our slices. The circumplanetary disc (CPD) rotates prograde (bottom panel) and emits a low-density bipolar outflow (top panel). The bipolar wind appears driven by steep pressure gradients near CPD surface layers.}
    \label{fig:contours}
\end{figure*}

\subsection{Overview and Diagnostics}\label{sec:over}

Simulations typically play out as follows. They begin with a period of gravito-turbulence which intensifies as $\beta = \Omega t_{\rm cool}$ (or the analogous parameter $b$ in the optically thin runs) decreases and cooling accelerates. 
At some point during the cooling the disc fragments, usually when $\beta$ reaches its prescribed minimum of 2.
The vertical disc velocities generated during the pre-fragmentation, gravito-turbulent phase will lead to non-zero fragment obliquities. \citet{Shi2014} found that such velocities approach but remain below the sound speed for most of the disc mass (see their figure 7). We confirm that the disc just prior to fragmentation is still relatively vertically thin --- see the progression of thin colored lines in Fig.~\ref{fig:zprofile} from high to low $\beta$.
Snapshots of the evolution of our
high-resolution, $\beta_0 = 20$ run 
are shown in Fig.~\ref{fig:proj}. 

Depending on the run, between one and four
fragments initially form. Multiple fragments subsequently merge. Occasionally a fragment ``spins off'' another when one of its extended tidal tails collapses under its own self-gravity. Fragments are formally located by searching for local
minima in the self-gravitational potential $\Phi$. The outer boundary of a fragment
is the equipotential surface whose
largest distance $R_{\rm max}$ from the enclosed
local potential minimum 
coincides with the Hill 
radius $R_{\rm Hill} = (GM_{\rm frag}/3\Omega^2)^{1/3}$,
where $M_{\rm frag}$ is the mass
contained within this outermost
equipotential (Fig.~\ref{fig:contours}, left panel). The radii $R_{\rm max}$
and $R_{\rm Hill}$ are computed
iteratively until $R_{\rm max} = R_{\rm Hill}$
to within 1 simulation cell width.
We confirm that
the Bondi radius $R_{\text{Bondi}} = 2GM_{\rm frag}/\langle c_{\rm s}^2 \rangle_{\rho}$,
where $\langle ... \rangle_{\rho}$
denotes a density-weighted average over
the fragment,
is always larger
than $R_{\rm Hill}$; fragment sizes are
limited by stellar tides and
not by thermal pressure, as appears clear
from the two-armed tidal tails extending
out to the fragment boundary (Figs.~\ref{fig:proj} and \ref{fig:contours}). The fragment center of mass
is computed over all cells within the fragment boundary, which we reiterate is not assumed spherical
but conforms to a level surface of $\Phi$. In practice, the center of mass
differs negligibly from the potential minimum.

The inertial-frame spin
angular momentum vector of a fragment
is computed as

\begin{align} \label{eq:J}
\bm{J} = \sum_i m_i \bm{R}_i \times \left( \bm{V}_i  + \Omega \bm{\hat{z}}
  \times \bm{R}_i  \right)
\end{align}
where $m_i$, $\bm{R}_i$, and $\bm{V_i}$ 
are the mass, displacement vector, and velocity vector
of the $i$th cell, measured in the shearing-box frame relative to the center of mass, and 
the sum is over all cells within
the fragment boundary. The term $+\Omega
\bm{\hat{z}}  \times \bm{R}_i$, 
where $\Omega = 1$ is the frame rotation frequency, corrects for the
apparent retrograde ``precession'' of the fragment's spin axis seen in
the shearing-box frame (e.g. Nesvorny et al.~2019, but note that their
equation 8 has a sign error for the correction term). 
This precession, which occurs with angular frequency $-\Omega\bm{\hat{z}}$, is clearly evident in movies of our simulations (see Fig.~\ref{fig:proj} for two
snapshots at two different precession phases). It is
a purely kinematic effect
that arises because the shearing-box frame
rotates with angular frequency $+\Omega \bm{\hat{z}}$. 
In practice, removing the frame rotation to recover the
inertial-frame $\bm{J}$ is a small correction.

The obliquity $\Psi$ of a fragment is the angle between the spin $\bm{J}$ and the orbit normal. As noted in the Introduction, a local shearing box cannot
track the evolution of the orbital angular momentum vector (there is no unique orbital radius in a local calculation), and so we have no choice but to
assume the orbit normal always points in the $\bm{\hat{z}}$
direction, as it does by construction at the beginning
of our simulations. Thus we compute the obliquity as
\begin{equation} \label{eq:obliquity}
    \cos{\Psi} = \frac{\bm{J} \cdot \bm{\hat{z}}}{|\bm{J}|} \,.
\end{equation}

\subsection{Obliquities}

\begin{figure*}
    \centering
    \includegraphics[width=\textwidth]{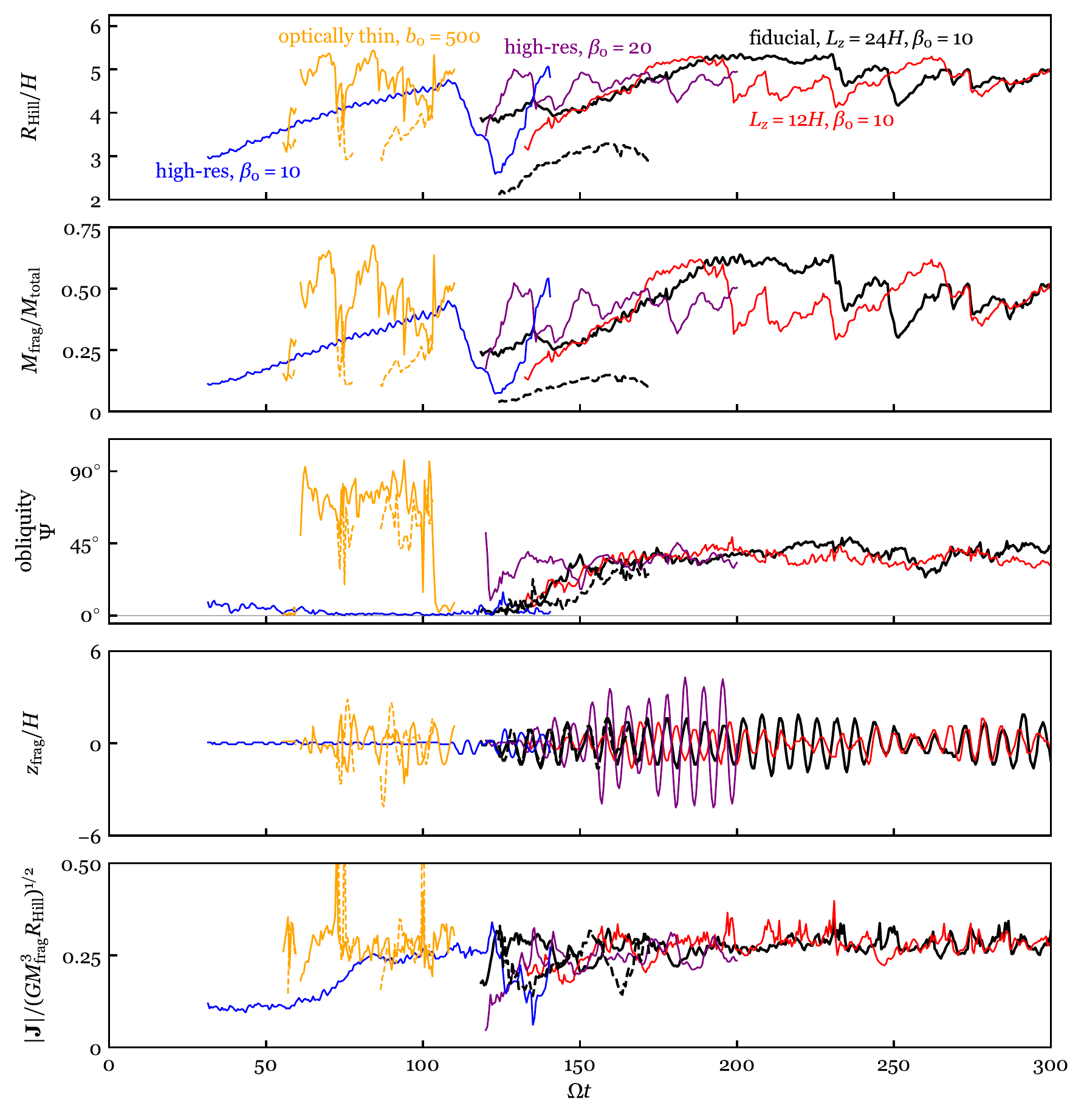}
    \caption{Time evolution of fragment properties. From top to bottom: Hill radius $R_{\rm Hill}$, fragment mass $M_{\rm frag}$ relative to the total mass in the simulation box, obliquity $\Psi$, height $z_{\rm frag}$ of the fragment center of mass, and total spin angular momentum $|\bm{J}|$ normalized to a rough measure of the spin angular momentum needed for break-up. Different colors refer to different simulations as annotated. Different linestyles (solid vs. dashed) refer to different fragments within the same simulation. For our fiducial run (black), two fragments appear at $\Omega t \sim 125$ and eventually merge into one at $\Omega t \sim 170$. The optically thin, $b_0 = 500$ run (yellow) is especially dynamic: two fragments first appear at $\Omega t \sim 60$ and merge soon thereafter, 
    and the resultant object periodically spins off and re-accretes a lower-mass companion. 
    Large obliquities can arise either at the time a fragment forms and accretes (black, purple, red)
    or when fragments collide (yellow). 
    The disc in the high-resolution, $\beta_0 = 10$ run (blue) is initially fast-cooling and fragments before turbulence has had a chance to develop; the resultant single fragment has low obliquity. Spikes
    in the bottommost panel are due either to close encounters between fragments which rapidly alter spin angular momenta (yellow), or to sudden decreases and increases in fragment mass which change the denominator $\propto M_{\rm frag}^{5/3}$ (blue). See text for more details.
    }
    \label{fig:frag_quantities}
\end{figure*}

Fig.~\ref{fig:frag_quantities} plots fragment obliquities $\Psi$ from our various simulations.
In our fiducial $\beta$-cooling run (black curve: $L_z = 24H$, resolution = 4 cells per $H$, initial $\beta_0 = 10$), the disc initially spawns two fragments having small obliquities. As the fragments interact with each other and leftover circumstellar (CS) gas, the obliquities increase 
from a combination of gravitational forcing and direct accretion of mass and angular momentum.
Eventually the fragments merge into a single object having $\Psi \simeq 25$--45$^\circ$. 

In other $\beta$-cooling runs, the disc fragments into
a single object whose obliquity appears determined
from the spin angular momentum of accreted gas and
not from gravitational forcing by ambient material.
Obliquities on the order of 1 radian are produced in both the small box run (red curve: $L_z = 12H$), and the high-resolution run with slower initial cooling (dashed blue: $8$ cells per $H$, $\beta_0=20$).

Fragment masses $M_{\rm frag}$ toward the end of the aforementioned simulations range from 30\% to 65\% of the total (conserved) box mass $M_{\rm total}$, fluctuating as fragments repeatedly collide with disc filaments. 
Sometimes the colliding filament arises from the
fragment itself: because of our shearing-periodic radial boundaries, a fragment's tidal arm can extend past one radial boundary to re-emerge from the other boundary and swipe the original fragment.
This self-swiping is not necessarily unphysical; the arm that collides with the fragment could be interpreted as originating from another fragment outside the simulation domain. In any case, 
the fluctuations in mass are confined to the 
periphery of the fragment and do not appear to much
affect the obliquity (see also the next section
\ref{sec:struc} on fragment internal structure).

In the high-resolution $\beta$-cooling run with faster initial cooling (solid blue: $8$ cells per $H$, $\beta_0=10$), $\Psi$ is small, $\lesssim 10^\circ$, a consequence of the fragment
forming early, before gravito-turbulent vertical motions fully develop (see also Fig.~\ref{fig:other_proj}). This case underscores how the approach to fragmentation matters: large obliquities more easily arise in discs that are already turbulent before they collapse. The obliquity remains small despite the fragment losing $\sim$90\% of its mass in an apparent self-swipe; the fragment's
small obliquity and close confinement to the disc midplane (see the panel in Fig.~\ref{fig:frag_quantities} plotting $z_{\rm frag}$, the height of the fragment center of mass) make it especially susceptible to self-swiping. The dynamics here are essentially of 2D midplane flows. During the mass loss and subsequent re-accretion, the obliquity varies between zero and $10^\circ$.

Discs in the optically thin $b$-experiments fragment relatively easily, before vertical turbulent motions develop. Hot, overpressured gas created at the intersection of converging flows and colliding filaments does not stay hot for long, as hotter gas cools proportionately faster according to our cooling prescription (\ref{eq:opt}). In a sense, the equation
of state is soft, and so when disc gas compresses in one direction, it is more prone to gravitationally collapse than to flow in
another direction. 
Initially in these optically thin runs, midplane gas collapses into more numerous and lower-mass fragments, all having small obliquities, as compared to our $\beta$-cooling models. In the $b_0 = 600$ run, between one and four low-obliquity fragments are present near the midplane at any given time, with the number changing from mergers and tidal spin-offs (Fig.~\ref{fig:other_proj}, middle column). 
In the $b_0 = 500$ run, two fragments initially form with low obliquities, and soon collide to form an object with a near-90$^\circ$ obliquity. The subsequent evolution is chaotic (Figs.~\ref{fig:frag_quantities} and \ref{fig:other_proj}), with multiple spin-offs and mergers leading eventually to a single low-obliquity object.

\subsection{Fragment Internal Structure}\label{sec:struc}
Fig.~\ref{fig:rprof} profiles
the radial structure of a fragment (the same one displayed in Fig.~\ref{fig:proj}),
plotting the mass, scalar spin angular momentum,
and obliquity of material enclosed within
equipotentials of varying size $R$ (the largest
distance between a given equipotential surface and the
center of mass). Roughly half of the total fragment mass is contained within an ``inner core'' of size $R \simeq 0.2 R_{\rm Hill}$. This inner core contains 20--25\% of the total spin angular momentum, and its mass profile $M_{\rm frag}(R)$ appears to scale linearly with $R$, suggesting the core is supported against self-gravity in large part by thermal pressure (an isothermal sphere in hydrostatic equilibrium has an enclosed mass that scales linearly with radius). 

The other half of the fragment mass is located in the ``outer circumplanetary disc (CPD)'' which extends from $R \simeq 0.2R_{\rm Hill}$ to $R_{\rm max} = R_{\rm Hill}$. About 80\% of the total spin angular momentum (whose magnitude $|\bm{J}|$ is comparable to the break-up value; see Fig.~\ref{fig:frag_quantities}) is carried by the outer CPD. There is a slight warp across the outer CPD as $\Psi(R)$ decreases by several degrees from small to large $R$; see also the volume rendering in Fig.~\ref{fig:volume}.

\subsection{Vertical Dynamics}\label{sec:vert}
Disc vertical velocities arising from gravito-turbulence result in fragments that not only rotate obliquely but also have vertical center-of-mass motions. From Fig.~\ref{fig:frag_quantities}, in those runs where $\Psi \sim 1$ rad, the vertical position of a fragment $z_{\rm frag}$ oscillates with a typical amplitude of $\sim$1--$2H$;
in one high-resolution run, vertical excursions extend up to $\sim$$4H$. In the runs where $\Psi < 10^\circ$, vertical oscillations are correspondingly small, $|z_{\rm frag}| \ll H$.

Vertical oscillations imply that fragments have non-zero orbital inclinations---unsurprisingly, given their formation within a turbulent disc. The orbital inclinations cannot be read directly from our local shearing-box simulations which have no well-defined orbital radius $a$. However, if we assume an aspect ratio of $H/a \sim 0.1$ typical of protoplanetary discs, then vertical oscillation amplitudes of 1--4$H$ imply orbital inclinations of 0.1--0.4 rad $\sim$ 6--20$^\circ$. 
In making this estimate we have
assumed that the midplane pressure scale height $|dz/d\ln P| \sim 0.1 a$ of an actual disc can be approximated by our code-based $H \equiv \Sigma_0/(2\rho_0)$, where the latter is defined as part of our hydrostatic initial condition. This identification is good if real discs have density profiles resembling our assumed initial condition within a few pressure scale heights of the midplane   where most of the disc mass 
resides. Fig.~\ref{fig:zprofile} lends support to this assumption by showing that fast-cooling, gravito-turbulent, pre-fragmentation discs have vertical density profiles within $z = \pm 2H$ resembling those of our hydrostatic initial condition.

As obliquity is the angular separation
between an object's spin axis and orbit normal,
it obviously depends on orbital inclination. A shortcoming of our formula (\ref{eq:obliquity}) for
$\Psi$ imposed by the shearing box that it does not account for
non-zero inclination. We expect, however,
the error to be small to the extent that our estimated inclinations of 6--20$^\circ$ are smaller than our measured 45--$90^\circ$ spin axis tilts. 
Furthermore, our estimated inclinations should not bias our measured values for $\Psi$ upward or downward, as the phase of the vertical oscillations
in $z_{\rm frag}$ does not correlate with the ``precession'' phase of the spin axis (see section \ref{sec:over}).

Fragments gravitationally stir leftover circumstellar (CS) disc material strongly. For the high-resolution, $\beta_0=20$ simulation plotted in Fig.~\ref{fig:zprofile}, after the fragment forms, the ambient CS disc thickens dramatically; 
gas densities at the vertical boundaries of our
simulation domain remain within a factor of $\sim$10 of the midplane density. The CS disc midplane itself is forced out of the $z=0$ plane; Fig.~\ref{fig:zprofile} shows the maxima of the post-fragmentation vertical density profiles (dashed lines) varying between $z\simeq \pm 2H$. The gravity of the fragment is responsible for this forcing of CS disc gas. The fragment mass is an order-unity fraction of the original CS disc mass (Fig.~\ref{fig:frag_quantities}),
and its surface escape velocity $\sim$$\sqrt{2 GM_{\rm frag}/R_{\rm Hill}}$ is 
$\sim$$2.5 \times$ larger than the CS disc sound speed
evaluated at the midplane just prior to fragmentation.

\begin{figure}
    \centering
    \includegraphics[width=\columnwidth]{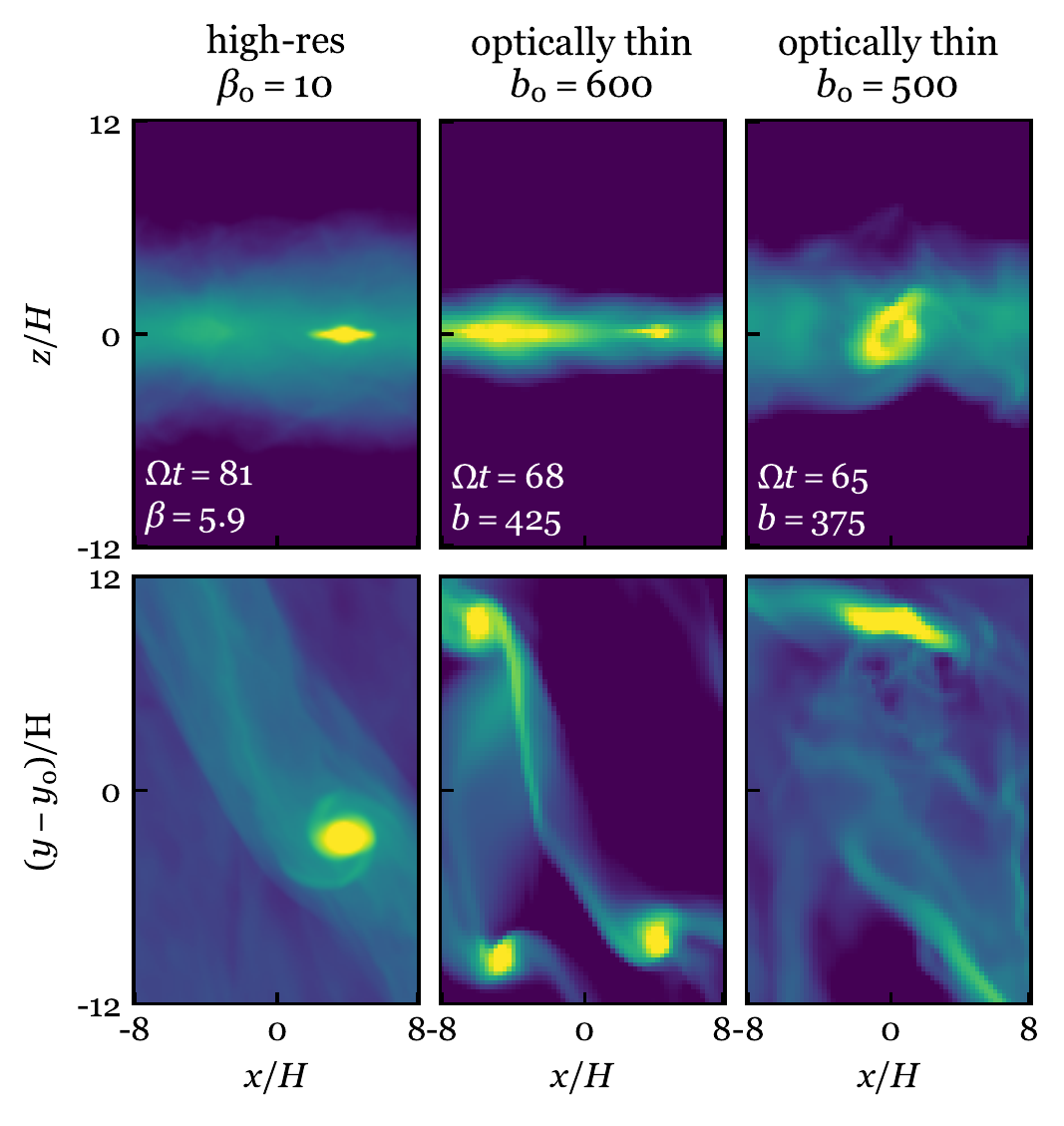}
    \caption{More column density snapshots like those in  Fig.~\ref{fig:proj}, using the same color scale. Obliquities are small in the high-resolution $\beta_0=10$ run and in the optically thin $b_0 = 600$ run (for the latter we show only 3 of the 4 fragments that form initially). In the optically thin $b_0=500$ run, several fragments momentarily coalesce into a rotating ring having an obliquity of nearly 90$^\circ$.}
    \label{fig:other_proj}
\end{figure}

\begin{figure}
    \centering
    \includegraphics[width=\columnwidth]{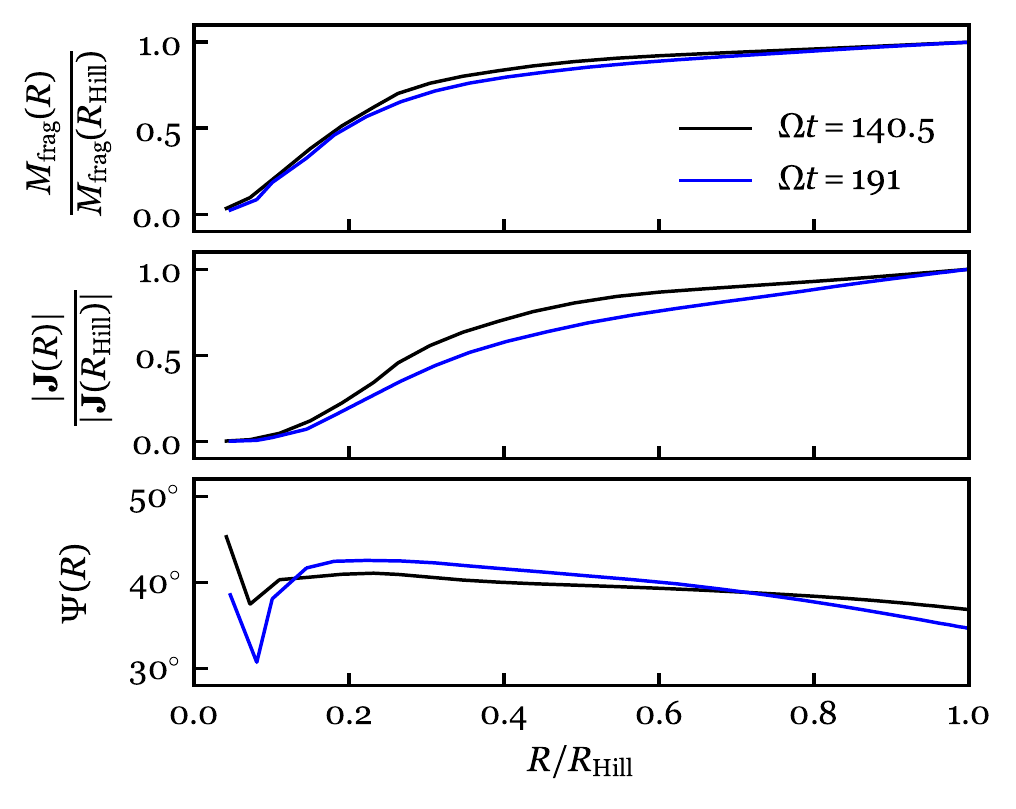}
    \caption{Fragment internal structure, plotting the mass, spin angular momentum, and obliquity of material enclosed with an equipotential of running size $R$.
    More than half of the total fragment mass and about half of the total spin angular momentum are contained within $R \sim 0.3 R_{\rm Hill}$. Data are shown for the high-resolution, $\beta_0=20$ run at two times as indicated. There are about 32 simulation cells spanning $R_{\rm Hill} \simeq 4H$.}
    \label{fig:rprof}
\end{figure}

\begin{figure}
    \centering
    \includegraphics[width=0.9\columnwidth]{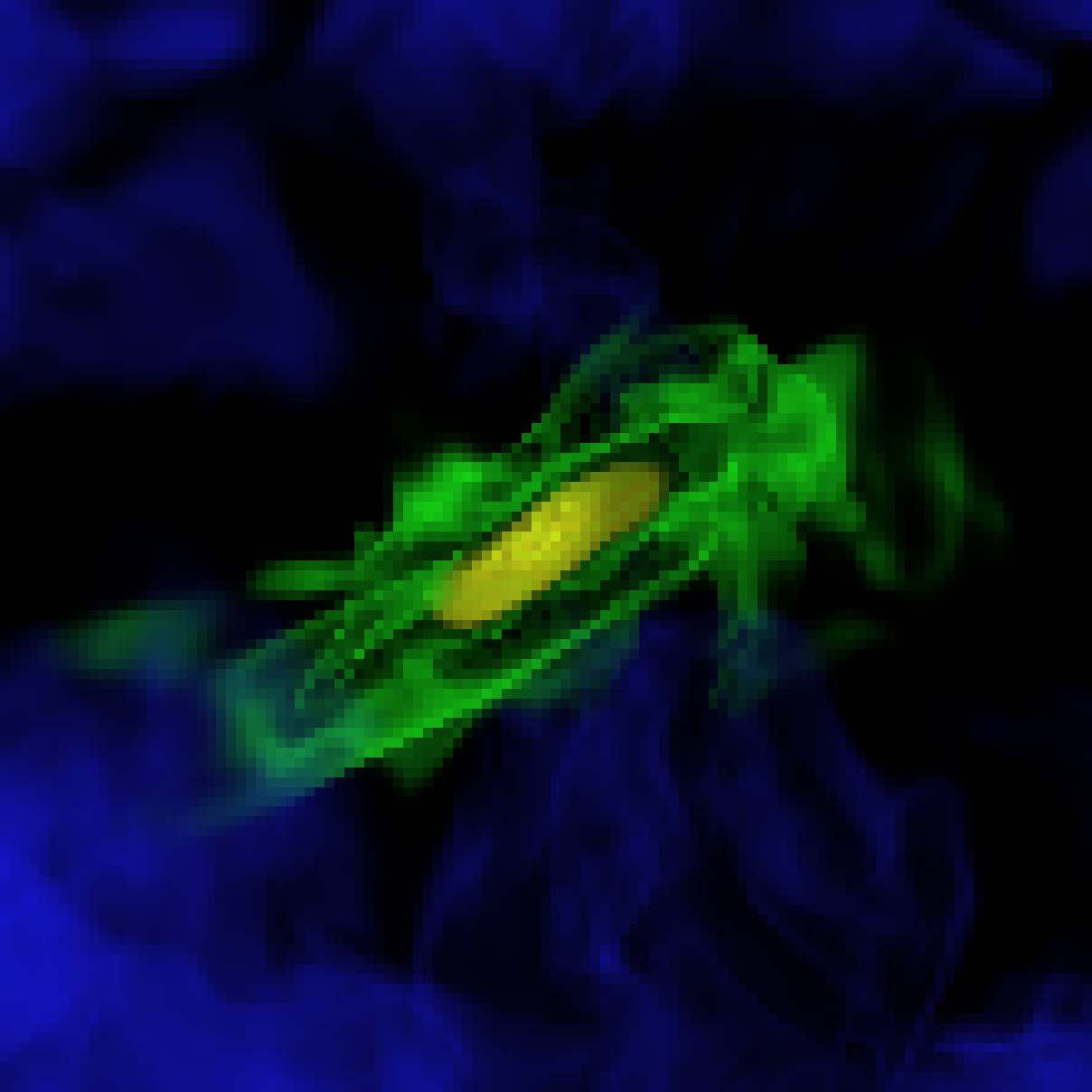}
    \caption{Volume rendering of the fragment in our high resolution, $\beta_0 = 20$ run at $\Omega t = 191$. Gas with $\num{1.8e3} \geq \rho/\rho_0 \geq 10^2$ is colored yellow, $10^2 > \rho/\rho_0 \geq 1$ is green, and $1 > \rho/\rho_0 \geq 10^{-2}$ is blue. The yellow inner core is $\sim$$1H$ in diameter.}
    \label{fig:volume}
\end{figure}

We emphasize that the most extreme thickening of the disc, when an order-unity fraction of the disc mass is thrown to heights exceeding several $H$, occurs only after the fragment forms; this 
violent puffing-up is a consequence and not a cause of fragmentation, and does not affect the fragment obliquity. 
For example, the high-resolution $\beta_0 = 20$ run which yields high obliquity (Fig.~\ref{fig:proj}) and the high-resolution $\beta_0=10$ run which produces low obliquity (Fig.~\ref{fig:other_proj}) exhibit   post-fragmentation disc thickening of similar magnitude.  Obliquities derive instead from whatever vertical velocities are present in the disc pre-fragmentation. Near the midplane, within $z = \pm 2H$, these velocities approach but remain less than the sound speed (\citealt{Shi2014}, their figure 7). Fig.~\ref{fig:zprofile} shows
that the disc just prior to fragmentation is much thinner than the disc after fragmentation --- contrast the thin yellow lines for $\beta \approx 2$ with the thick dashed curves. Just before it fragments, the disc between $z=\pm 2H$ still contains the lion's share of the mass, and its vertical density profile is similar to that of the hydrostatic initial condition. Non-zero obliquity follows less from puffing up --- there isn’t much puffing before fragmentation --- and more from the disc acquiring non-zero turbulent vertical velocities because of self-gravity.

\section{Summary and Discussion} \label{sec:sum}
Brown dwarf and super-Jupiter companions
to stars are thought to form by gravitational instability in circumstellar discs (e.g.  \citealt{Nielsen2019}).
Self-gravitating discs are 
turbulent, containing a ``swirling hotch-potch of pieces of spiral arms'' \citep{Goldreich1965,Gammie2001,Shi2014}. We have shown by direct numerical
simulation that gravito-turbulence can lead to gravitationally collapsed fragments that spin with order-unity obliquities. Initial obliquities are determined by
the spin angular momentum of accreted material; a nascent fragment spins in the same direction as the turbulent whorl from which it was born. In our simulations, this primordial spin can point as much as $45^\circ$ from the orbit normal. Subsequent close encounters, collisions, and mergers of fragments can increase or decrease obliquities, up to $\sim$$90^\circ$ or down to near-zero.

The caveat to these results is that
they were obtained in a shearing box for
which stellocentric radius and by extension 
orbital angular momentum and orbital inclination
are not defined. Thus we have assumed
in computing obliquities that orbit normals
are fixed in the direction they were
pointing at the start of our simulations,
parallel to the box's vertical dimension.
At the same time we have seen both the fragments
and the leftover circumstellar disc gas 
oscillate
vertically in our simulations, implying that
non-zero orbital inclinations are excited in the course of disc
evolution. These vertical motions are expected from 
gravito-turbulence and gravitational forcing
by fragments. We have measured the vertical oscillations to have an
amplitude comparable to the vertical disc
scale height $H$. Insofar as real-life protoplanetary discs have height-to-radius ratios of 
 $H/a \sim 0.1$, we estimate that the corresponding orbital inclinations
are on the order of 0.1 rad, somewhat smaller
than the order-1 spin tilts measured
in several of our runs. 
Furthermore,
these vertical oscillations appear randomly
phased with respect to spin axis
orientations, implying that our
turbulence-induced obliquities
are not systematically reduced by non-zero
orbital inclinations. 
Thus our conclusion that order-unity obliquities can result from gravitational instability
seems secure, pending global 3D simulations
that can simultaneously track
spin and orbit. Note further that in all our runs,
fragments move radially, sometimes across the entire simulation domain spanning $16H$; these radial motions are only to be expected from the background turbulence, and suggest that objects formed by gravitational instability should have large orbital eccentricities.

The approach to fragmentation, i.e. the 
exact manner in which Toomre's $Q$ and the gas
cooling time $t_{\rm cool}$ drop to the critical
values needed for gravitational collapse, matters for obliquity. If the descent to instability 
is fast (lasting less than ten or so orbital periods), or if overdensities cool so efficiently that gas pressure cannot compete against self-gravity (in other words if the equation of state is too soft), then
vertical turbulent motions may not develop in the disc before gas collapses, more-or-less symmetrically, into a low-obliquity fragment. 
The evolution to fragmentation depends
on the radiation-thermodynamics of how gas
cools, how mass is transported across disc
radius, and ongoing accretion from the disc's natal envelope. These issues are 
beyond the scope of our study. 
\citet[][their section 6]{Kratter2010} discuss a fragmentation scenario whereby a disc is fed at too high a rate by an external infalling envelope for disc spiral density waves to torque the mass away. The local surface density then builds until $Q$ falls to order-unity; meanwhile, the cooling criterion is assumed to be readily satisfied (their figure 1; see also \citealt{Tobin2016}, their extended data figure 4). The disc becomes
  gravito-turbulent, and potentially spawns an obliquely rotating
  fragment, if the timescale for the local disk to grown in mass and $Q$ to approach unity is at least as
  long as the local orbital time; this seems a realistic prospect. Note that this scenario, which assumes the cooling criterion is always satisfied while $Q$ decreases, is the converse of what we implemented in our code for numerical convenience; the underlying point is that disc evolutionary timescales can be slow enough to trigger turbulence before fragmentation.

Gravito-turbulence sets an object's initial obliquity. A litany
of other effects may subsequently 
alter its value, either by changing the spin axis or the orbital axis. 
These include, e.g., torques
from the host star that can either
damp or excite the inclination of a
circumplanetary disc relative to the
circumstellar orbit plane \citep{Lubow2000,Martin2021}; 
collisions between protoplanets (e.g. \citealt{Li2020}; \citealt{LiLai2020}); changes to the orbit plane
from ongoing circumstellar disc accretion \citep{Tremaine1991}; and spin-orbit resonances brought about by migration \citep{Millholland2019} or
planet-planet scattering \citep{Hong2021,Li2021}. 
Planetary magnetic fields are thought to be critical in regulating the spin rates of proto-gas-giants and thereby enabling their contraction to observed sizes \citep{Ginzburg2020}; 
magnetic torques may change
the spin direction as well (e.g. \citealt{Spalding2015}).

Obliquities of objects having masses upwards of $\sim$10 Jupiter masses are
beginning to be constrained observationally.
The first brown
dwarfs/super-Jupiters 
to be examined for their spin-orbit
angles are 2M0122b \citep{Bryan2020} and HD 106906b (Bryan et
al. 2021, in press).
Both companions orbit their host stars at the
large, $> 50$ AU distances conducive to disc fragmentation, and both
evince, modulo large uncertainties, order-unity obliquities. In the case of the HD 106906 system, it has been established with high confidence that the companion is viewed nearly pole-on (Bryan et al. 2021, in press) while
a debris disc orbiting the same host is viewed nearly edge-on \citep{Kalas2015,Lagrange2016,Nguyen2021}.
Disc gravitational instability
is also relevant for the formation of binary stars.
Whether stellar-mass secondaries
should have large obliquities
is unclear; it may be that such companions accrete so much of the parent 
disc that turbulent velocity fluctuations in the accreted material average to zero. Observational errors on the degree of
spin-orbit alignment in solar-type binaries are currently too large to draw conclusions \citep{Justesen2020}.


\section*{Acknowledgements}
We thank Ji-Ming Shi for providing a copy of his code, and Marta Bryan, Dan Fabrycky, Sivan Ginzburg, Greg Laughlin, Dong Lai, Gongjie Li, Chris Matzner, and Zhaohuan Zhu 
for helpful discussions. Kaitlin Kratter provided a thoughtful referee's report which improved the content of this paper. The simulations were run on the
Savio computer cluster provided by the Berkeley Research Computing program, which is supported by the UC Berkeley Chancellor, Vice Chancellor for Research, and Chief Information Officer. This research made use of the open source project {\tt yt} (\href{https://yt-project.org}{yt-project.org}; \citealt{Turk2011}), and was supported by NASA grant NNX15AD95G/NEXSS.

\section*{Data Availability} 
Movies of the simulations may be accessed at \href{https://michaeljennings11.github.io/pages/gi}{https://michaeljennings11.github.io/pages/gi}.




\bibliographystyle{mnras}
\bibliography{GI} 



\bsp	
\label{lastpage}
\end{document}